\documentclass{article}

\usepackage{arxiv}

\usepackage[utf8]{inputenc} % allow utf-8 input
\usepackage[T1]{fontenc}    % use 8-bit T1 fonts
\usepackage{hyperref}       % hyperlinks
\usepackage{url}            % simple URL typesetting
\usepackage{booktabs}       % professional-quality tables
\usepackage{amsfonts}       % blackboard math symbols
\usepackage{nicefrac}       % compact symbols for 1/2, etc.
\usepackage{microtype}      % microtypography
\usepackage{lipsum}
\usepackage{graphicx}
\usepackage{xcolor}

\title{AMII: Adaptive Multimodal Inter-personal and Intra-personal Model for Adapted Behavior Synthesis}

\author{Jieyeon Woo \\
	ISIR - Sorbonne University \\
	France \\
	\texttt{woo@isir.upmc.fr} \\
	\And
	Mireille Fares \\
	ISIR - Sorbonne University \\
	France \\
	\texttt{fares@isir.upmc.fr} \\
	\And
	Catherine Pelachaud \\
	CNRS - ISIR - Sorbonne University \\
	France \\
	\texttt{pelachaud@isir.upmc.fr} \\
	\And
	Catherine Achard \\
	ISIR - Sorbonne University \\
	France \\
	\texttt{achard@isir.upmc.fr} \\
}

\begin{document}

\definecolor{darkgreen}{rgb}{0.0, 0.7, 0.0}
\definecolor{lightred}{rgb}{1., 0.44, 0.37}

\maketitle
\begin{abstract}
Socially Interactive Agents (\textit{SIA}s) are physical or virtual embodied agents that display similar behavior as human multimodal behavior. Modeling \textit{SIA}s’ non-verbal behavior, such as \textit{speech} and \textit{facial gestures}, has always been a challenging task, given that a \textit{SIA} can take the role of a \textit{speaker} or a \textit{listener}. A \textit{SIA} must emit appropriate behavior adapted to its own \textit{speech}, its previous behaviors (\textit{intra-personal}), and the \textit{User}’s behaviors (\textit{inter-personal}) for both roles. We propose \textit{AMII}, a novel approach to synthesize \textit{adaptive} \textit{facial gestures} for \textit{SIA}s while interacting with \textit{Users} and acting interchangeably as a \textit{speaker} or as a \textit{listener}. \textit{AMII} is characterized by \textit{modality memory encoding schema} - where \textit{modality} corresponds to either \textit{speech} or \textit{facial gestures} - and makes use of \textit{attention mechanisms} to capture the \textit{intra-personal} and \textit{inter-personal} relationships. We validate our approach by conducting objective evaluations and comparing it with the state-of-the-art approaches.
\end{abstract}


\section{Introduction}
\textit{Socially Interactive Agents} (\textit{SIA}s) are embodied agents emitting human-like non-verbal behavior such as \textit{facial}, \textit{body}, and \textit{hand} gesturing during \textit{speech}~\cite{cassell2000human}. In many applications such as medical therapy~\cite{shidara2022automatic} and educational assistance~\cite{kim2017embodied}, they can carry out interactive and natural conversations with humans (\textit{Users}) by sending and receiving multimodal non-verbal signals~\cite{cappella1991mutual, burgoon2011nonverbal} in addition to the verbal message. To render \textit{SIA}-\textit{User} communication effective and engaging, \textit{SIA}s must play active roles of \textit{speakers} and \textit{listeners}. For instance, in successful and engaging human-human conversations, \textit{speaker}'s and \textit{listener}'s behaviors are constantly coordinated and adapted to each other~\cite{burgoon1995interpersonal}, through the perception and response of the multimodal emitted signals such as linguistic~\cite{pickering2004toward}, facial~\cite{burgoon2011nonverbal}, and body~\cite{knapp2013nonverbal} features.
The behavior adaptation can be: (1) \textit{intra-personal} which is between the multimodal signals of a single person or (2) \textit{inter-personal} which is between \textit{listener}'s and \textit{speaker}'s signals adapting to the other's verbal and non-verbal behavior.
This adaptation is \textit{continuous}, \textit{reciprocal} and \textit{dynamical} and is referred to as \textit{reciprocal adaptation}~\cite{woo2023reciprocal}. 
The display of such adaptation capability can enable \textit{SIA}s to be perceived as \textit{social} and \textit{engaging}~\cite{biancardi2021adaptation}. 

Various works have generated \textit{SIA}s' non-verbal behavior focusing on only the modeling of \textit{intra-personal} relationship for a single person~\cite{alexanderson2020style,bhattacharya2021text2gestures,fares2023zero}.
Other works looked into capturing multimodal \textit{inter-personal} information in dyadic settings. 
In the work of Feng et al.~\cite{feng2017learn2smile} and Dermouche et al.~\cite{dermouche2019generative}, \textit{SIA} \textit{facial gestures} are synthesized based on past gestures of both \textit{SIA} and \textit{User} without considering the existing relation with \textit{audio modality}~\cite{pell2005prosody, yehia2002linking}, and do not ensure the \textit{motion continuity}. Grafsgaard et al.~\cite{grafsgaard2018generative} synthesize interlocutor's \textit{gestures} based on  both the interlocutors' \textit{audio} and their \textit{facial modalities}. 
In the work of Jonell et al.~\cite{jonell2020let}, \textit{SIA}'s \textit{facial gestures} are generated based on \textit{SIA}'s \textit{speech} and the \textit{User}'s \textit{speech} and \textit{facial gestures}. However, these models~\cite{grafsgaard2018generative, jonell2020let} are prone to produce \textit{non-continuous gestures}.
The works presented in Woo et al.~\cite{woo2021creating, woo2023asap} and Ng et al.~\cite{ng2022learning} ensure \textit{SIA}'s \textit{behavior continuity} by employing autoregressive online inference while modeling \textit{SIA}'s and \textit{User}'s multimodal features. Only the \textit{listener}'s behavior is modeled in \cite{ng2022learning}. \cite{woo2021creating, woo2023asap} and all the aforementioned models for dyadic setting do not explicitly model the \textit{intra-personal} relationship.

Our overall aim is to create a \textit{social} and \textit{engaging} \textit{SIA} by modeling its \textit{behavior adaptation}. Our hypothesis is that \textit{behavior adaptation} is captured by \textit{intra-personal} and \textit{inter-personal} synchronizations.
We propose \textbf{\textit{Adaptive Multimodal Inter-personal and Intra-personal (AMII)}} model, a novel method to synthesize \textit{adaptive} and \textit{engaging} \textit{facial} gesturing for \textit{SIA}s. We explicitly model the \textit{intra-personal} relationship by encoding the prior emitted multimodal signals (\textit{modality memory}) while ensuring \textit{motion continuity}. We model the \textit{inter-personal} relationship to generate \textit{SIA} behavior for both roles of \textit{speaker} and \textit{listener}. We explore the best way to capture the \textit{reciprocal adaptation} to generate \textit{adaptive} non-verbal \textit{SIA} behavior within a dyadic setting. 
\textit{Intra-personal} and \textit{inter-personal} relations are learned through \textit{attention mechanisms}. 
We demonstrate that our \textit{AMII} model outperforms the state-of-the-art models, in terms of both \textit{behavior appropriateness} and \textit{reciprocal adaptation resemblance}, by conducting objective evaluations. We perform ablation studies and show the influence of \textit{AMII}'s key encoders.

Our contributions are as follows:
\begin{itemize}
    \item We propose \textit{AMII}, an approach to capture the \textit{reciprocal adaptation} of \textit{SIA} behaviors.
    \item We explicitly model the \textit{intra-personal} and \textit{inter-personal} relationships by encoding prior emitted multimodal signals via our \textit{modality memory} encoders.
\end{itemize}

The paper is organized as follows. Section~\ref{section:model} describe the proposed \textit{AMII} model architecture. The dataset is presented in Section~\ref{section:dataset} and we detail the training regime in Section~\ref{section:training}. Then, we report our experiments in Section~\ref{section:exp} and discuss our results in Section~\ref{section:discussion}. We finally conclude and talk about our future works in Section~\ref{section:conc}.

\section{Adaptive Multimodal Inter-personal and Intra-personal (\textit{AMII}) Model Architecture}
\label{section:model}

\begin{figure*}[ht!]
  \centering
  \includegraphics[width=0.9\linewidth]{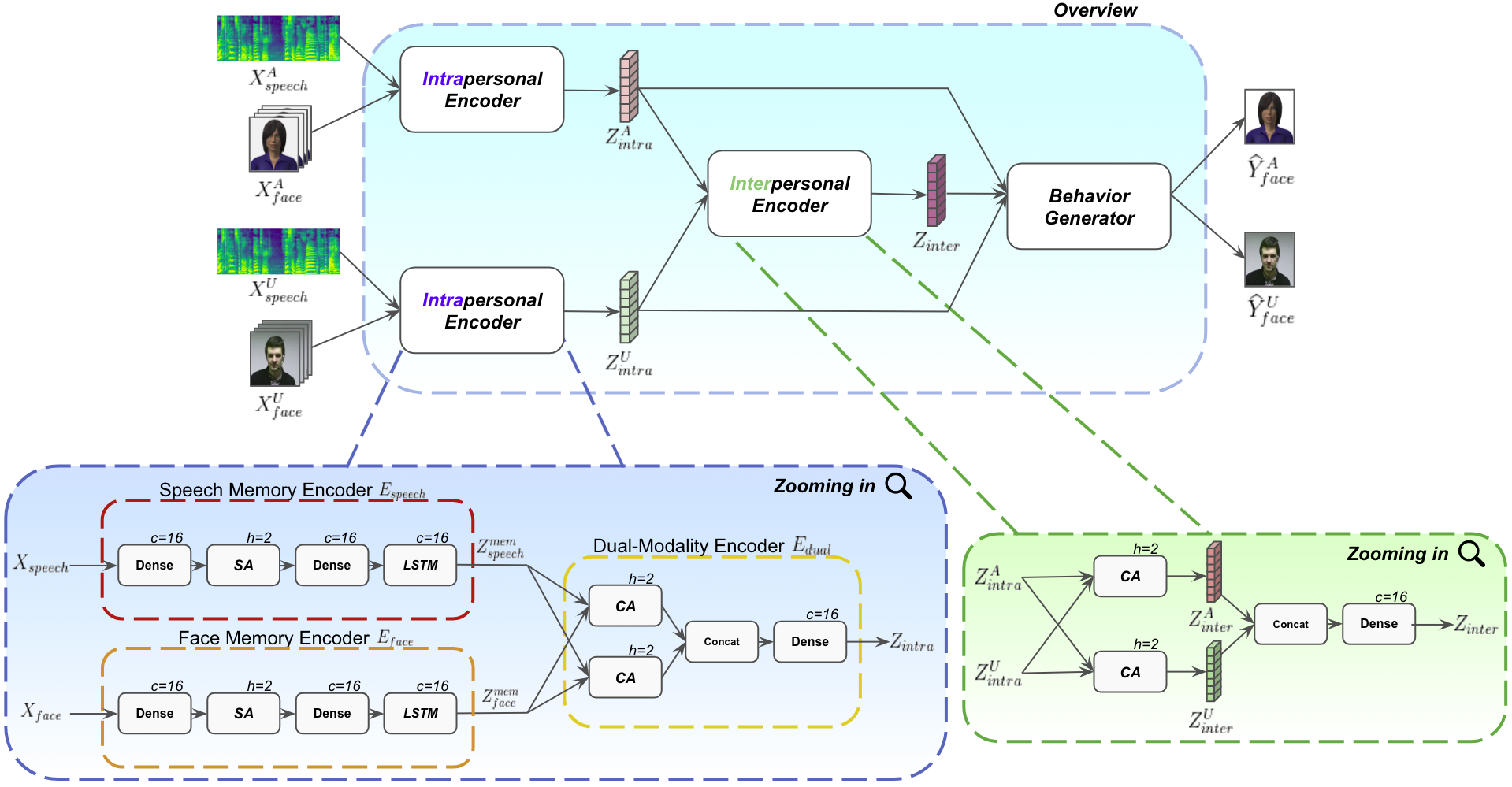}
  \caption{\textbf{\textit{AMII}} (\textbf{A}daptive \textbf{M}ultimodal \textbf{I}ntra-personal and \textbf{I}nter-personal) model architecture. The intra-personal encoder ($E_{intra}$) takes the \textit{speech} $X_{speech}$ and the \textit{facial gestures} $X_{face}$ of the previous 100 frames of either the \textit{SIA} ($A$) or the \textit{User} ($U$) to encode the corresponding \textit{intra-personal} relationship $Z_{intra}$. The inter-personal encoder ($E_{inter}$) learns from \textit{intra-personal} relationships $Z_{intra}^{A}$ and $Z_{intra}^{U}$ to encode the \textit{inter-personal} relationship between them $Z_{inter}$. The behavior generator ($G_{face}$) takes $Z_{intra}^{A}$, $Z_{intra}^{U}$, and $Z_{inter}$ to generate the sequence of \textit{facial gestures} for the next frame t+1 $\widehat{Y}_{face}^{A}$ and $\widehat{Y}_{face}^{U}$. At \textit{training time}, \textit{AMII} is trained with human-human ($U_{1}$-$U_{2}$) interactions ($U_{1}$ for $A$ and $U_{2}$ for $U$) and predicts both of humans' facial gestures ($\widehat{Y}_{face}^{U_{1}}$ and $\widehat{Y}_{face}^{U_{2}}$). At inference time, \textit{AMII} renders the facial gestures of $A$ and $U$. To infer the next $A$'s behavior, we feed back the predicted $A$'s behavior and the ground truth of $U$.}
  \label{fig:AMII}
\end{figure*}

We focus on generating \textit{adaptive} non-verbal \textit{SIA} behavior as both a \textit{speaker} and \textit{listener}. For this purpose, we propose an approach trained on real human-human interactions, to learn human-human \textit{inter-personal} and \textit{intra-personal} relations for \textit{SIA} and simulate our predictions on a \textit{SIA}. We propose a new model architecture, the Adaptive Multimodal Inter-personal and Intra-personal model (\textit{AMII}), to synthesize \textit{adaptive} and \textit{engaging} \textit{facial gestures} for \textit{SIA}s. It takes as input \textit{speech} and \textit{facial gestures} of both \textit{SIA} ($A$) and \textit{User} ($U$), corresponding to their past behavior, and predicts \textit{SIA}'s and \textit{User}'s \textit{facial gestures} at the next time step.

\textbf{\textit{Facial gestures}} are represented by:
\begin{itemize}
    \item \textbf{Gaze movements: }represented by $G_{x}$ and $G_{y}$ which are the gaze angles w.r.t. the  $x$ and $y$ axis.
    \item \textbf{Head movements: }represented by $R_{x}$, $R_{y}$, and $R_{z}$ which are the Euler head rotations w.r.t. the $x$, $y$ and $z$ axis.
    \item \textbf{Facial movements: }represented by facial Action Units (AUs)~\cite{ekman1976measuring} which are facial muscle movements defined by the Facial Action Coding Systems (FACS)~\cite{ekman1978facial}. We use \textit{AU1} (inner brow raiser), \textit{AU2} (outer brow raiser), \textit{AU4} (brow lowerer), \textit{AU6} (cheek raiser), and \textit{AU12} (lip corner puller for smile).
\end{itemize}

\textbf{\textit{Speech and facial gestures}} are highly tied together~\cite{pell2005prosody, yehia2002linking}. With this relation, we decide to use \textit{speech} information to drive \textit{SIA}'s \textit{facial gestures}. \textbf{\textit{Speech features}} are listed as follows:
\begin{itemize}
    \item \textbf{Pitch: }represented by the fundamental frequency $f_{0}$, which is the main prosodic feature correlated with \textit{facial gestures}~\cite{bolinger1958theory}. 
    \item \textbf{Loudness: }speech intensity from the auditory spectra.
    \item \textbf{Voicing probability: }
    speech presence probability expressed as a probability score in the range of $0$ to $1$.
    \item \textbf{Mel-Frequency-Cepstral Coefficients (MFCC): }represented by $13$ \textit{MFCC} features ($0$-$12$).    
\end{itemize}

\textbf{\textit{AMII}} model operates as follows. It takes as input the $100$ past frames ($t-99:t$), where $t$ is the current frame, of the:
\begin{enumerate}
    \item \textit{Speech features} of $A$ ($X_{speech}^{A}$) and those of $U$ ($X_{speech}^{U}$),
    \item \textit{Facial features} of $A$ ($X_{face}^{A}$) and those of $U$ ($X_{face}^{U}$).
\end{enumerate}
For each prediction of the next frame ($t+1$), the model predicts:
\begin{enumerate}
    \item $A$'s \textit{facial gestures} ($\widehat{Y}_{face}^{A}$), 
    \item $U$'s \textit{facial gestures} ($\widehat{Y}_{face}^{U}$).
\end{enumerate}

\textit{AMII} consists of three main components, as illustrated in Figure~\ref{fig:AMII}. The first component is the intra-personal encoder $E_{intra}$, which explicitly encodes the \textit{intra-personal} relations via a \textit{modality memory schema}. This schema consists of encoding each \textit{modality} - \textit{speech features} and \textit{facial features} - corresponding to the past 100 frames. The second component is the inter-personal encoder $E_{inter}$, which encodes the \textit{inter-personal} relations by applying cross-attentions between \textit{A}'s and \textit{U}'s features' embeddings. The last component is the \textit{behavior generator} which generates \textit{A}'s and \textit{U}'s \textit{facial gestures} of the next frame. These components are detailed in the following.
\subsection{Intra-personal Encoder (\texorpdfstring{$E_{intra}$}{Lg})}
As shown in Figure~\ref{fig:AMII}, $E_{intra}$ takes as input $X_{speech}$ and $X_{face}$ of either \textit{A} or \textit{U} and generates the corresponding intra-personal embedding $Z_{intra}$. It consists of two sub-encoders. The first is the modality memory encoder ($E_{speech}$ or $E_{face}$). The second is the dual-modality encoder $E_{dual}$.

\paragraph{Modality Memory Encoder ($E_{speech}$ or $E_{face}$)}Both $E_{speech}$ and $E_{face}$ takes its corresponding \textit{modality} - $X_{speech}$ or $X_{face}$ respectively - as input and renders the modality memory embedding $Z_{speech}^{mem}$ and $Z_{face}^{mem}$ representing the past $100$ frames. 
Each corresponding modality memory encoder firstly learns the \textit{modality specific} information by applying \textit{self-attention}, with a $h=2$ (where $h$ is the head size), preceded and followed by dense layers ($E_{d}$) with $c=16$ (where $c$ is the cell size). Then, it embeds the memory sequence of the chosen \textit{modality} via a LSTM layer ($E_{m}$) with $c=16$, as depicted in Figure~\ref{fig:AMII}. 
It takes $X_{mod}$ - where $mod$ represents either \textit{speech} or \textit{facial gestures} - and outputs $Z_{mod}^{mem}$, and can be expressed as:
\begin{equation}
Z_{mod}^{mem}=E_{m}\left(E_{d}\left(SA\left(E_{d}\left(X_{mod}\right)\right)\right)\right)
\end{equation}
where $SA$($\cdot$) denotes self-attention layer.

\paragraph{Dual-modality Encoder ($E_{dual}$)}$E_{dual}$ captures the relationship between the multimodal signals by applying \textit{cross-attention mechanisms} on the corresponding modalities: $CA^{speech}$ and $CA^{face}$ with $h=2$ followed by $E_{d}$ with $c=16$, as shown in Figure~\ref{fig:AMII}. $CA^{speech}$ has a query $Q$ equals to $Z_{speech}^{mem}$ with key $K$ and value $V$ equal to $Z_{face}^{mem}$. $CA^{face}$ has a query $Q$ equals to $Z_{face}^{mem}$ with key $K$ and value $V$ equal to $Z_{speech}^{mem}$.
$E_{dual}$ takes $Z_{speech}^{mem}$ and $Z_{face}^{mem}$ as inputs and generates $Z_{intra}$. It can be written as:
\begin{equation}
Z_{intra}=
E_{d}\left(\left[ \right.\right.
CA_{speech}\left(Q_{speech},K_{face},V_{face}\right), \\
CA_{face}\left(Q_{face},K_{speech},V_{speech}\right)
\left.\left. \right]\right)
\end{equation}
where $CA$($Q,K,V$) denotes cross-attention layer and $[\cdot]$ denotes concatenation layer.
\subsection{Inter-personal Encoder (\texorpdfstring{$E_{inter}$}{Lg})}
As illustrated in Figure~\ref{fig:AMII}, $E_{inter}$ takes as input $Z_{intra}^{A}$ and $Z_{intra}^{U}$, which are the \textit{intra-personal} representations of $A$ and $U$ respectively. It renders $Z_{inter}$, a representation of \textit{inter-personal} relation between $A$ and $U$. 
$E_{inter}$ applies \textit{cross-attention mechanisms} on the both \textit{intra-personal} representations: $CA^{A}$ and $CA^{U}$ with $h=2$ followed by $E_{d}$ with $c=16$. $CA^{A}$ has a query $Q$ equals to $Z_{intra}^{A}$ with key $K$ and value $V$ equal to $Z_{intra}^{U}$. $CA^{U}$ has a query $Q$ equals to $Z_{intra}^{U}$ with key $K$ and value $V$ equal to $Z_{intra}^{A}$. It can be written as:
\begin{equation}
Z_{inter}=
E_{d}\left(\left[CA^{A}\left(Q^{A},K^{U},V^{U}\right), CA^{U}\left(Q^{U},K^{A},V^{A}\right)\right]\right)
\end{equation}
where $CA$($Q,K,V$) denotes cross-attention layer and $[\cdot]$ denotes concatenation layer.
\subsection{Behavior Generator (\texorpdfstring{$G_{face}$}{Lg})}
$G_{face}$ takes as input the:
\begin{enumerate}
    \item $A$'s \textit{intra-personal} representation ($Z_{intra}^{A}$),
    \item $U$'s \textit{intra-personal} representation ($Z_{intra}^{U}$),
    \item \textit{Inter-personal} representation of $A$ and $U$ ($Z_{inter}$).
\end{enumerate}
It generates the corresponding \textit{facial gestures} $\widehat{Y}_{face}^{A}$ and  $\widehat{Y}_{face}^{U}$ by decoding with a dense layer ($D_{d}$) with $c=20$, as depicted in Figure~\ref{fig:AMII}.
The final outputs $\widehat{Y}_{face}^{P}$ can be written as:
\begin{equation}
\widehat{Y}_{face}^{P}=D_{d}\left(Z_{intra}^{P},Z_{inter}\right)
\end{equation}
where $P$ represents either \textit{A} or \textit{U}. At training time, \textit{AMII} synthesizes $\widehat{Y}_{face}^{U}$ to better learn \textit{inter-personal} and \textit{intra-personal} relations. $\widehat{Y}_{face}^{U}$ is disregarded at inference time since the aim is to predict only $A$.
\subsection{Training and Inference Modes}
We train our model on real human-human ($U_{1}$-$U_{2}$) interactions. Our model learns to synthesize adapted gestures of $U_{1}$ and $U_{2}$.
During inference, \textit{AMII} synthesizes the behavior of $A$ and $U$. $\widehat{Y}_{face}^{A}$ is inferred using the previous prediction of $A$ and the ground truth of $U$. We apply \textit{adaptive online prediction} to generate \textit{continuous} \textit{A}'s behavior in an \textit{autoregressive} fashion.
\section{NoXi Dataset}
\label{section:dataset}
For our experiments, we choose the French \textit{NoXi} dataset~\cite{cafaro2017noxi} consisting of $21$ human-human dyadic interactions performed by $28$ participants with a total duration of $7h22min$.
We extract non-verbal behavior features for each time-step: openSMILE~\cite{eyben2010opensmile} for \textit{speech} and OpenFace~\cite{baltruvsaitis2016openface} for \textit{facial gestures}.
Data preprocessing - median filter and linear interpolation - is applied on both extracted features and they are adjusted to $25fps$.
We split our dataset into $3$ sets: training ($70\%$), validation ($10\%$), and test ($20\%$). The test set does not include data of \textit{speakers} and \textit{listeners} that are seen during training. The aim is to test \textit{AMII}'s capacity to extrapolate on new unseen \textit{speakers} and \textit{listeners} and therefore its capability to generalize.
\section{Training Regime}
\label{section:training}

To train our model, we use the Mean Squared Error (MSE) as our loss function and the Adam optimizer~\cite{kingma2014adam} with Cyclical Learning Rate (CLR)~\cite{smith2017cyclical} (\textit{triangular} learning rate policy, \textit{base\_lr} of $1e-7$, \textit{max\_lr} of $1e-3$, and step size factor of $10$). 
The training was done for $300$ epochs (with an average runtime of $125h$) on a $2.2GHz$ Intel Xeon Linux server with NVIDIA GeForce GTX TITAN X and $64GB$ RAM with a batch size of $32$. The best set of hyperparameters is chosen after manual optimization based on the validation set. 
\section{Experiments}
\label{section:exp}
To assess our model, we conduct objective evaluation to check its performance against the state-of-the-art approaches, which we select as our baselines, and to verify the effectiveness of each \textit{AMII}'s key components through ablation studies.

\subsection{Objective Metrics}
We want to assess whether the generated behavior is \textit{appropriate} (\textit{intra-personal}) and \textit{reciprocally adaptive} (\textit{inter-personal}). To do so, we employ the metrics used in previous works~\cite{feng2017learn2smile, dermouche2019generative, grafsgaard2018generative, woo2021creating, woo2023asap, ng2022learning}.
We measure \textit{behavior appropriateness} of $A$'s predictions ($\widehat{A}$) against its ground truth (GT) behavior ($A$). The metrics are as follows:
\begin{itemize}
    \item \textbf{MAE} and \textbf{RMSE}/\textbf{L2}: distance between the predictions and GT to measure the generated error, which are used in \cite{feng2017learn2smile, dermouche2019generative, woo2021creating, woo2023asap, ng2022learning}.
    \item \textbf{Kolmogorov-Smirnov two-sample test (KS test)}~\cite{massey1951kolmogorov}: density probability difference between \textit{A}'s corresponding predictions and GT to check for the distribution similarities between them, which is used in \cite{woo2023asap}.
\end{itemize}

For \textit{reciprocal adaptation resemblance}, we measure the resemblance between $\widehat{A}$ and $U$'s GT data ($U$). We choose to assess only smile (\textit{AU12}) as it is a key socio-emotional signal~\cite{knapp2013nonverbal}. The metrics are as follows:
\begin{itemize}
    \item \textbf{Time lagged cross correlation coefficient (TLCC)}~\cite{boker2002windowed}: linear relationship invariant to speed, which is used to quantify \textit{global synchrony} like PCC. TLCC is computed in chunks of $8sec$ with a time lag of $2sec$ as in \cite{ng2022learning}.
    \item \textbf{DTW}~\cite{muller2007dynamic}: proximity/resemblance check of $\widehat{A} \&U$ against that of the GT interaction (between two humans) to evaluate \textit{reciprocal adaptation}. DTW is computed in chunks of $1min$ with a stride of $30sec$ as in \cite{woo2023asap}.
    \item \textbf{Synchrony (Sync) and Entrainment Loop (EL)}~\cite{woo2023reciprocal}: synchrony and entrainment loop measures proposed in \cite{woo2023reciprocal} to evaluate the \textit{reciprocal adaptation} between \textit{$\widehat{A}$} and \textit{U}.
\end{itemize}
Lower values denote better performance for MAE, RMSE, and KS test. For the \textit{resemblance metrics} (PCC, TLCC, DTW, Sync, and EL), the closer the value of the metric is to the GT, the better the model performs in generating \textit{adaptive} \textit{A}'s behaviors.

\subsection{Baseline Models and Ablation Studies}
We consider the following \textit{baseline models (base)}:
\begin{itemize}
    \item \textbf{IL-LSTM}~\cite{dermouche2019generative}: models only the \textit{inter-personal} based on a single modality of \textit{facial gestures} of \textit{A} and \textit{U},
    \item \textbf{Symmetrized IL-LSTM with online LSTM (sym-IL-LSTM)}~\cite{woo2021creating} and \textbf{ASAP}~\cite{woo2023asap}: model the \textit{inter-personal} relation of \textit{A} and \textit{U}, \textit{multimodality} of \textit{speech} and \textit{facial gestures}, and assure \textit{motion continuity}.
\end{itemize}

To check for the effectiveness and influence of each of \textit{AMII}'s key encoders, we conducted additional ablation studies. We performed the ablations of:
\begin{itemize}
    \item Modality Memory Encoder $E_{m}$ ($noE_{m}$),
    \item Dual-modality Encoder $E_{dual}$ ($noE_{dual}$),
    \item Inter-personal Encoder $E_{inter}$ ($noE_{inter}$).
\end{itemize}

\section{Results and Discussion}
\label{section:discussion}

\begin{table*}[ht!]
  \caption{Objective evaluation of AMII against the baselines along with ablations using the selected metrics. GT denotes ground truth interaction. Best results are highlighted in \textbf{bold}. $\Delta_{base}$ represents the change in performance over the best performing baseline approach of each metric. $\Delta_{base}$ entries in \textcolor{darkgreen}{green} when AMII outperforms best baseline, in \textcolor{lightred}{red} when it is not the case.}
  \label{tab:eval_obj}
  \centering
  \begin{tabular}{ l | c c c c c c c c}
  \toprule
  & \multicolumn{1}{c}{MAE}
  & \multicolumn{1}{c}{RMSE}
  & \multicolumn{1}{c}{KS test}
  & \multicolumn{1}{c}{TLCC}
  & \multicolumn{1}{c}{DTW}
  & \multicolumn{1}{c}{Sync}
  & \multicolumn{1}{c}{EL}\\  
  \midrule
  GT 
    & 
    &   
    & 
    & 0.334
    & 1317.5
    & 132.4
    & 1172.3\\
  \hline
  IL-LSTM~\cite{dermouche2019generative}
    & 0.304
    & 0.415
    & 0.329  
    & 0.343
    & 1216.2
    & 45.3
    & 323.3\\
  sym-IL-LSTM~\cite{woo2021creating}  
    & 0.180
    & 0.227
    & 0.284  
    & \textbf{0.335}
    & 1281.3
    & 33.3
    & 232.3\\
  ASAP~\cite{woo2023asap}
    & 0.185
    & 0.254
    & \textbf{0.282} 
    & 0.317
    & 1399.3
    & 142.0
    & 1890.5\\
  \hline
  AMII-$noE_{m}$ (ours)
    & 0.099
    & 0.132
    & 0.515   
    & 0.271
    & 1228.4
    & 79.1
    & 603.8\\
  AMII-$noE_{dual}$ (ours)
    & 0.143
    & 0.186          
    & 0.396   
    & 0.300
    & 1352.6
    & 262.3
    & 2255.8\\
  AMII-$noE_{inter}$ (ours)
    & 0.136
    & 0.178          
    & 0.406 
    & 0.261
    & 1127.8
    & 82.3
    & 586.3\\
  AMII (ours)
    & \textbf{0.156}
    & \textbf{0.197} 
    & 0.437  
    & 0.291
    & \textbf{1319.6}
    & \textbf{137.4}
    & \textbf{989.0}\\
  \hline
  $\Delta_{base}$
    & \textbf{\textcolor{darkgreen}{$\downarrow$ 0.024}}
    & \textbf{\textcolor{darkgreen}{$\downarrow$ 0.030}}
    & \textcolor{lightred}{$\uparrow$ 0.155}
    & \textcolor{lightred}{$\uparrow$ 0.042}
    & \textbf{\textcolor{darkgreen}{$\downarrow$ 79.7}}
    & \textbf{\textcolor{darkgreen}{$\downarrow$ 4.6}}
    & \textbf{\textcolor{darkgreen}{$\downarrow$ 534.9}}\\
  \bottomrule
  \end{tabular}
\end{table*}

The evaluation results are listed in Table~\ref{tab:eval_obj}. $\Delta_{base}$ represents the change in performance over the best performing baseline approach for each metric.

\paragraph{Comparing with Baselines}We remark that \textit{AMII} outperforms the \textit{baselines} in terms of \textit{behavior appropriateness}. This is reflected through low errors of MAE and RMSE represented by $\Delta_{base}$ ($\downarrow$ $0.024$ and $\downarrow$ $0.020$ respectively). For the density distribution, via the KS test, we observe that \textit{AMII} performs comparatively less than the \textit{baselines} indicating that \textit{AMII} possesses the least similar density distribution compared to that of the GT. In detail, ASAP performs the best in terms of having the most similar density w.r.t. GT ($0.282$) and AMII the worst ($0.437$) with $\Delta_{base}$ of $\uparrow$ $0.155$.
This low performance of \textit{AMII} does not imply that it generates wrong \textit{SIA} behavior but that it has either a smaller or a wider range of behavior variety than that of the GT. The focus of this study is not to produce a variety of behaviors but to generate \textit{SIA} behaviors that are \textit{adaptive} to its interlocutor. Thus, this weak performance of the KS test metric is not critical for our aim. 
Moreover, \textit{AMII} performs the best in terms of the \textit{reciprocal adaptation resemblance} metrics as seen in the Table~\ref{tab:eval_obj}. DTW, synchrony, and entrainment loop of $\widehat{A} \&U$ show that \textit{AMII} resembles the GT the most with $\Delta_{base}$ of $\downarrow$ $79.7$, $\downarrow$ $4.6$, and $\downarrow$ $534.9$ respectively.  
With TLCC, we remark that the sym-IL-LSTM is the closest to the GT while \textit{AMII} is the farthest one with $\Delta_{base}$ of $\uparrow$ $0.042$. 
As DTW considers the variation of sequence length while being invariant to speed unlike TLCC, it represents better the \textit{global correlation}. Thus, for the interpretation, we can put more emphasis on the DTW results compared to that of TLCC.
This comparative study shows that the inclusion of explicit modeling of \textit{intra-personal} relation leverages the quality of produced gestures in terms of both \textit{behavior appropriateness} and \textit{reciprocal adaptation resemblance}.

\paragraph{Ablation studies}The ablation of each of \textit{AMII} key encoder - $E_{m}$, $E_{dual}$, and $E_{inter}$ - results in the improvement of the \textit{reciprocal adaptation resemblance}. This was seen by an increase in DTW ($87.0$, $33.0$, and $187.6$ respectively), synchrony ($48.3$, $124.9$, and $45.1$ respectively), and entrainment loop resemblance ($385.2$, $900.2$, and $402.7$ respectively). 
TLCC shows that the insertion of $E_{m}$ improves the \textit{AMII} by $0.020$ along with $E_{inter}$ by $0.030$. However, $E_{dual}$ slightly deteriorates the performance by $0.009$. As in the \textit{baseline comparison study}, it is better to concentrate on the other \textit{reciprocal adaptation resemblance} metrics as DTW is a more dynamic measure of synchrony than TLCC.
However, this enhancement of \textit{reciprocal adaptation resemblance} is at the expense of lowering its \textit{behavior appropriateness} performance. This is observed via MAE ($0.057$, $0.013$, $0.020$ respectively) and RMSE ($0.065$, $0.011$, $0.019$ respectively). The same conclusion can be drawn by looking at KS test result. The fall of performance is seen for the additions of $E_{dual}$ ($0.041$) and $E_{inter}$ ($0.031$) while $E_{m}$ improves ($0.078$).
This compromise of losing \textit{behavior appropriateness} to gain an \textit{adaptive} one may be a good exchange. It is more valuable to generate \textit{SIA} behaviors with \textit{adaptation capacity} than to reproduce the same GT behavior. In fact, in a human-human interaction there could be multiple possible behaviors and generation timings facing a same interacting partner's behavior. This might vary depending on the various factors such as the context, situation, and interlocutor's personality and mood. 

We can conclude that it is important to model the \textit{intra-personal} relation with the encodings of $E_{m}$ and $E_{dual}$, and the \textit{inter-personal} relation with $E_{inter}$ to synthesize \textit{adaptive} non-verbal \textit{facial gestures} for both roles as \textit{speaker} and \textit{listener}.

\section{Conclusions and Future works}
\label{section:conc}
In this paper, we propose a new approach to generate \textit{adaptive} \textit{SIA} behavior as both \textit{speaker} and \textit{listener} by encoding the multimodality and \textit{intra-personal} and \textit{inter-personal} relationships. We conclude that \textit{AMII} model achieves state-of-the art performance notably in terms of \textit{reciprocal adaptation resemblance}. Our approach still has some limitations. We do not model the inter-personal relation memory which could improve our results. As the next step, we plan to add this component. 
We also plan to integrate and assess \textit{AMII} within a real-time \textit{SIA}-\textit{User} interaction. Our overall aim is to model \textit{SIA} that can \textit{socially engage} users in an interaction through its behaviors that capture \textit{intra-personal} and \textit{inter-personal} relations. In the next future, we will subjectively evaluate the dimensions of social attitudes and engagement to validate this hypothesis.

\bibliographystyle{unsrt}

\end{document}